\documentclass[article]{jdssv}

\usepackage{lmodern}
\usepackage{bm}
\usepackage{amsthm}
\usepackage{amsmath}
\usepackage{graphicx}
\usepackage{booktabs}
\usepackage[skip = 0.5 \baselineskip]{caption}

\author{Alessandro Gasparini \\ University of Leicester
  \And Tim P. Morris \\ MRC Clinical Trials Unit \\ at UCL
  \And Michael J. Crowther \\ University of Leicester
}
\Plainauthor{Alessandro Gasparini, Tim P. Morris, Michael J. Crowther}

\title{\pkg{INTEREST}: INteractive Tool for Exploring REsults from Simulation sTudies}
\Plaintitle{INTEREST: INteractive Tool for Exploring REsults from Simulation sTudies}
\Shorttitle{\pkg{INTEREST}}

\Abstract{
  Simulation studies allow us to explore the properties of statistical methods.
  They provide a powerful tool with a multiplicity of aims; among others: evaluating and comparing new or existing statistical methods, assessing violations of modelling assumptions, helping with the understanding of statistical concepts, and supporting the design of clinical trials.
  The increased availability of powerful computational tools and usable software has contributed to the rise of simulation studies in the current literature.
  However, simulation studies involve increasingly complex designs, making it difficult to provide all relevant results clearly.
  Dissemination of results plays a focal role in simulation studies: it can drive applied analysts to use methods that have been shown to perform well in their settings, guide researchers to develop new methods in a promising direction, and provide insights into less established methods.
  It is crucial that we can digest relevant results of simulation studies.
  Therefore, we developed \pkg{INTEREST}: an \emph{INteractive Tool for Exploring REsults from Simulation sTudies}.
  The tool has been developed using the \pkg{Shiny} framework in \proglang{R} and is available as a web app or as a standalone package.
  It requires uploading a tidy format dataset with the results of a simulation study in \proglang{R}, \proglang{Stata}, \proglang{SAS}, \proglang{SPSS}, or comma-separated format.
  A variety of performance measures are estimated automatically along with Monte Carlo standard errors; results and performance summaries are displayed both in tabular and graphical fashion, with a wide variety of available plots.
  Consequently, the reader can focus on simulation parameters and estimands of most interest.
  In conclusion, \pkg{INTEREST} can facilitate the investigation of results from simulation studies and supplement the reporting of results, allowing researchers to share detailed results from their simulations and readers to explore them freely.
}

\Plainkeywords{Simulation study, Monte Carlo, Visualisation, Reporting, R, Shiny, Replicability}

\Address{
  Alessandro Gasparini \\
  Biostatistics Research Group \\
  Department of Health Sciences \\
  University of Leicester \\
  George Davies Centre \\
  University Road \\
  Leicester \\
  LE1 7RH \\
  United Kingdom \\
  E-mail: \email{alessandro.gasparini@ki.se} \\
  \ \\
  Tim P. Morris \\
  MRC Clinical Trials Unit at UCL \\
  90 High Holborn \\
  London \\
  WC1V 6LJ \\
  United Kingdom \\
  \ \\
  Michael J. Crowther \\
  Biostatistics Research Group \\
  Department of Health Sciences \\
  University of Leicester \\
  George Davies Centre \\
  University Road \\
  Leicester \\
  LE1 7RH \\
  United Kingdom
}

\begin{document}

\section{Background}

Monte Carlo simulation studies are computer experiments based on generating pseudo-random observations from a known truth.
Statisticians usually mean \emph{Monte Carlo simulation study} when they say \emph{Simulation study}; throughout this article, we will just use \emph{simulation study} but this encapsulates Monte Carlo simulation studies.
Simulation studies have several applications and represent an invaluable tool for statistical research nowadays: in statistics, establishing properties of current methods is key to allow them to be used -- or not -- with confidence.
Sometimes it is not possible to derive exact analytical properties; for example, a large sample approximation may be possible, but evaluating the approximation in finite samples is required.
Approximations often require assumptions as well: what are the consequences of violating such assumptions?
Monte Carlo simulation studies come to the rescue and can help to answer these questions.
They also can help answer questions such as: is an estimator biased in a finite sample?
What are the consequences of model misspecification?
Do confidence intervals for a given parameter achieve the advertised/nominal level of coverage?
How does a newly developed method compare to an established one?
What is the power to detect a desired effect size under complex experimental settings and analysis methods?

\begin{figure}[h]
  \caption{Trend in published documents on simulation studies from 1960 onwards. The number of documents was identified on Scopus via the search key \code{TITLE-ABS-KEY ("simulation study") AND SUBJAREA (math)}, and the number of documents identified in 2019 is labelled on the plot.}
  \label{fig:trend}
  \centering
  \includegraphics[width = 0.95 \textwidth]{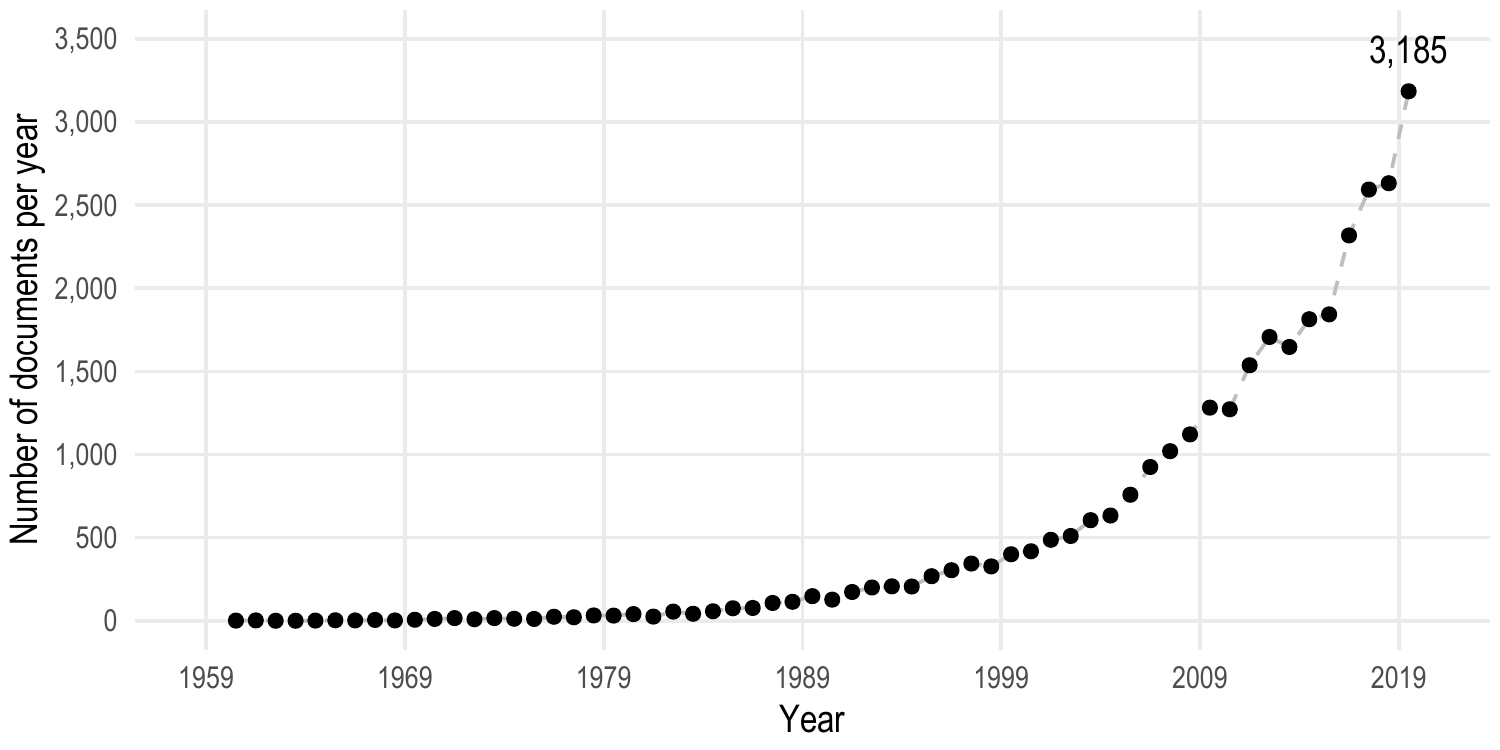}
\end{figure}

Simulation studies are being used increasingly in a wide variety of settings.
For instance, searching on the database of peer-reviewed research literature Scopus (\url{https://www.scopus.com}) with the query string \code{TITLE-ABS-KEY ("simulation study") AND SUBJAREA (math)} yields more than 30000 results with a 20-fold increase during the last 30 years, from 148 documents in 1989 to 3185 in 2019 (Figure \ref{fig:trend}).
The increased availability of powerful computational tools and ready-to-use software to researchers surely contributed to the rise of simulation studies in the current literature.

Despite the popularity of simulation studies, they are often poorly designed, analysed, and reported.
Morris \emph{et al}. reviewed 100 research articles published in Volume 34 of \emph{Statistics in Medicine} (2015) with at least one simulation study and found that information on data-generating mechanisms (DGMs), number of repetitions, software, and estimands were often lacking or poorly reported, making critical appraise and replication of published studies a difficult task \citep{morris_2019} .
Another aspect of simulation studies that is often poorly reported or not reported at all is the Monte Carlo error of estimated performance measures, defined as the standard error of estimated performance, owing to the fact that a finite number of repetitions are used and so performance is estimated with uncertainty.
Monte Carlo errors play an important role in understanding the role of chance in the results of simulation studies and have been showed to be severely underreported \citep{koehler_2009}.

The possibility of independently verifying results from scientific studies is a fundamental aspect of science \citep{laine_2007}; as a consequence, several reporting guidelines have emerged under the banner of the EQUATOR Network (\url{http://www.equator-network.org}) \citep{schulz_2010, vonelm_2007}.
Despite similar calls for harmonised reporting to allow for greater reproducibility in the area of computation science \citep{peng_2011} and several articles advocating for more rigour in specific aspects of simulation studies \citep{hoaglin_1975, hauck_1984, diazemparanza_2002, burton_2006, white_2010, smith_2011}, design and reporting guidelines for simulation studies are generally lacking in the statistical literature, with a few examples in the area of structural equation modelling \citep{bandalos_2012, boomsma_2013}.
Morris \emph{et al}. introduced the ADEMP framework (Aims, Data-generating mechanisms, Estimands, Methods, Performance measures) aiming to fill precisely that gap.
In the \emph{Reporting} section they compared the several ways of reporting results that they observed in their reviews, including results in text for small simulation studies, tabulating and plotting results, and even the nested-loop plot proposed by R\"{u}cker and Schwarzer for fully-factorial simulation studies with many data-generating mechanisms \citep{rucker_2014}.
They concluded by arguing that \emph{there is no correct way to present results, but we encourage careful thought to facilitate readability, considering the comparisons that need to be made}.

As outlined in Spiegelhalter \emph{et al}., there is little experimental evidence on how different types of visualisations are perceived \citep{spiegelhalter_2011}; despite that, they highlight the ease of improving understanding via interactive visualisations that can be adjusted by the user to best fit specific requirements.
The recent advent of tools such as Data-Driven Documents (\pkg{D\textsuperscript{3}}, or \pkg{D3.js}) \citep{bostock_2011} and \pkg{Shiny} \citep{shiny} has further facilitated the development of interactive visualisations.

The increased availability of powerful computational tools has not only contributed to a rise in the popularity of simulation studies, it has also allowed researchers to simulate an ever-growing number of data-generating mechanisms and include several estimands and methods to compare: up to \(4.2 \times 10 ^ {10}\), 32, and 33, respectively, in the aforementioned review \citep{morris_2019}.
With a large number of data-generating mechanisms, estimands, or methods, analysing and reporting the results of a simulation study becomes cumbersome: what results shall we focus on so as not to bewilder readers?
Which estimands and methods should we include in our tables and plots?
How should we plot or tabulate several data-generating mechanisms at once?

In an attempt to address these questions, we developed \pkg{INTEREST}, an \emph{INteractive Tool for Exploring REsults from Simulation sTudies}.
\pkg{INTEREST} is a browser-based interactive tool, and it requires first uploading a dataset with results from a simulation study; then, it estimates performance measures and it displays a variety of tables and plots automatically.
The user can focus on specific data-generating mechanisms, estimands, and methods: tables and plots are updated automatically.
This article will introduce the implementation details of \pkg{INTEREST} in the \emph{Implementation} section and the main features in the \emph{Results and discussion} section, where we will further discuss its relevance.
We also present a case study to motivate the use of INTEREST and illustrate its use in practice.
Finally, we conclude the manuscript with some ending remarks in the \emph{Conclusions} section.

\section{Implementation}

\pkg{INTEREST} was developed using the free statistical software \proglang{R} \citep{r} and the \proglang{R} package \pkg{Shiny} \citep{shiny}.
\pkg{Shiny} is an \proglang{R} package (and framework) that allows building interactive web apps straight from within \proglang{R}: the resulting applications can be hosted online, embedded in reports and dashboards, or just run as standalone apps.

The front-end of \pkg{INTEREST} has been built using the \pkg{shinydashboard} package \citep{shinydashboard}; \pkg{shinydashboard} is based upon \pkg{AdminLTE} (\url{https://adminlte.io/}), an open-source admin control panel built on top of the Bootstrap framework (Version 3.x) and released under the MIT license.

The back-end functionality of \pkg{INTEREST} is published as a standalone \proglang{R} package named \pkg{rsimsum} for easier long-term maintainability \citep{gasparini_2018}; \pkg{rsimsum} is freely available on the Comprehensive \proglang{R} Archive Network (CRAN) under the GNU General Public License Version 3 (\url{https://www.gnu.org/licenses/gpl-3.0}).

\pkg{INTEREST} is available as an online application and as a standalone version for offline use.
The online version is hosted at \url{https://interest.shinyapps.io/interest/}, and can be accessed via any web browser on any device (desktop computers, laptops, tablets, smartphones, etc.).
The standalone offline version can be obtained from GitHub (\url{https://github.com/ellessenne/interest}) and can be run on any desktop computer and laptop with a local instance of \proglang{R}; if required, \proglang{R} can be downloaded for free from the website of the \proglang{R} project \citep{r}.
INTEREST (as \pkg{rsimsum}) is published under the GNU General Public License Version 3.

\section{Results and discussion}

\begin{figure}
  \caption{Homepage of \pkg{INTEREST}. On the left, there is a navigation bar with sub-menus useful to tune the default behaviour of the app. On the right, the main window of \pkg{INTEREST}.}
  \label{fig:homepage}
  \centering
  \includegraphics[width = 0.95 \textwidth]{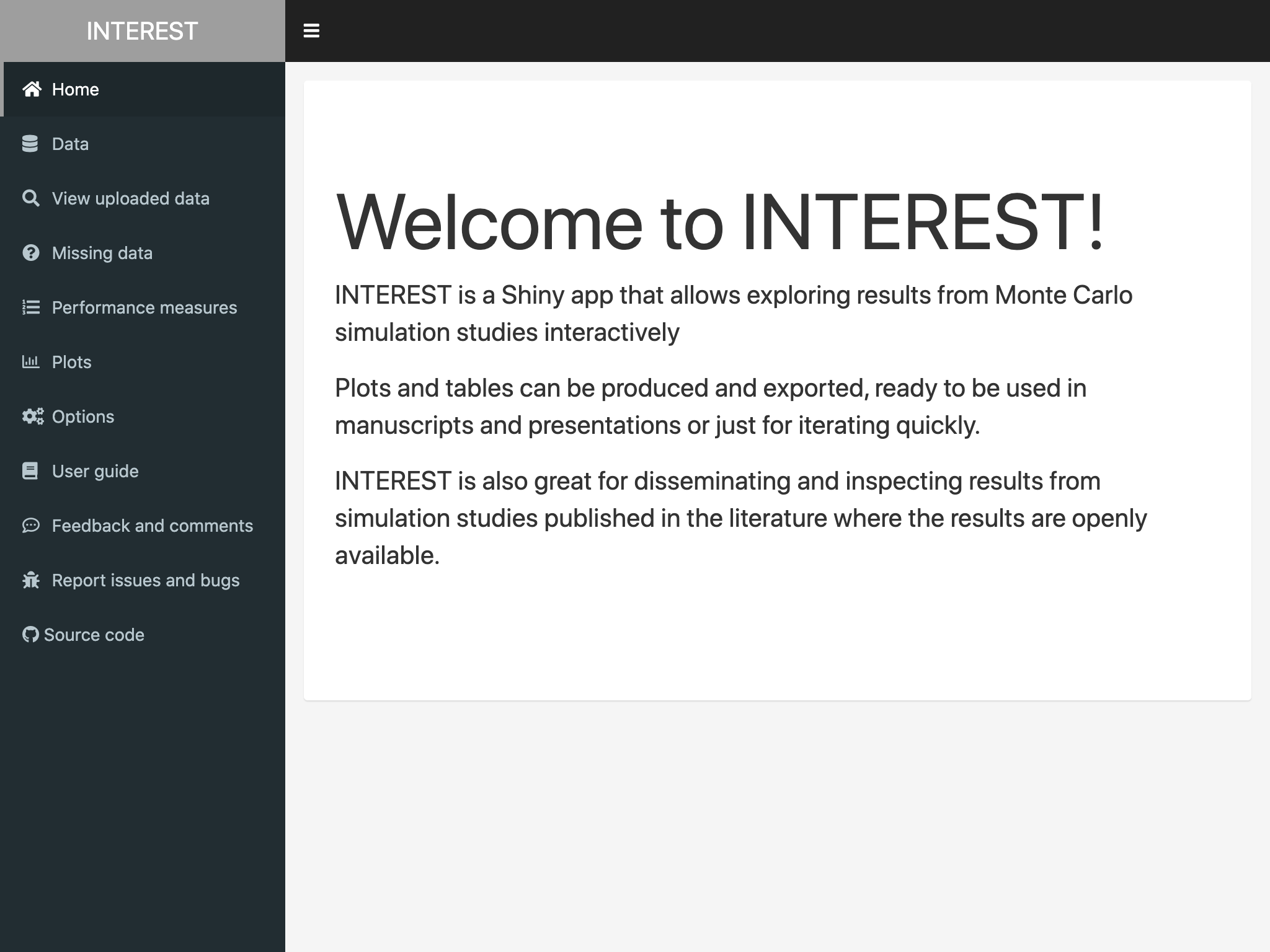}
\end{figure}

The main interface of \pkg{INTEREST} is presented in Figure \ref{fig:homepage}.
The interface is composed of a main area on the right and a navigation bar on the left; the navigation bar includes sub-menus for customising plots or modifying the default behaviour of \pkg{INTEREST}.
We now introduce and describe the functionality of the application.

\subsection{Data}

The use of \pkg{INTEREST} starts by providing a tidy dataset (also known as long format, with variables in columns and observations in rows \citep{wickham_2014}; an example of tidy data is included in Table \ref{tab:tidyex}) with results from a simulation study via the \emph{Data} tab from the side menu.
A dataset can be provided to \pkg{INTEREST} in three different ways:
\begin{enumerate}
    \item The user can upload a dataset.
    The uploaded file can be a comma-separated file (\code{.csv}), a \proglang{Stata} dataset (version 8-15, \code{.dta}), an \proglang{SPSS} dataset (\code{.sav}), a \proglang{SAS} dataset (\code{.sas7bdat}), or an \proglang{R} serialised object (\code{.rds}); the format will be inferred automatically from the extension of the uploaded file, and the auto-detection is case-insensitive.
    It is also possible to upload compressed files (ending in \code{.gz}, \code{.bz2}, \code{.xz}, or \code{.zip}) that are automatically decompressed.
    The maximum supported file size is 100MB;
    \item The user can provide a URL link to a dataset hosted elsewhere.
    All considerations relative to the file format from point (1) are also valid here;
    \item Finally, the user can paste a dataset (e.g. from Microsoft Excel) in a text box.
    The pasted data is assumed to be tab-separated.
\end{enumerate}
If users stored the results of their simulation study in a different format, we recommend using one of the readily available tools (e.g. the \code{pivot\_*} functions from the \pkg{tidyr} package in \proglang{R} or the \code{reshape} command in \proglang{Stata}) to reshape the data before uploading it to \pkg{INTEREST}.

\begin{table}
  \caption{Example of dataset in tidy format, with each row identifying a repetition for each combination of data-generating mechanism and analytical method.}
  \label{tab:tidyex}
  \centering
  \begin{tabular}{cccc}
    \hline
    Repetition & DGM        & Method     & Estimate \\
    \hline
    \code{1}  & \code{1} & \code{1} & \(\hat{\theta}_{1,1,1}\) \\
    \code{2}  & \code{1} & \code{1} & \(\hat{\theta}_{2,1,1}\) \\
    \code{3}  & \code{1} & \code{1} & \(\hat{\theta}_{3,1,1}\) \\
    \code{1}  & \code{2} & \code{1} & \(\hat{\theta}_{1,2,1}\) \\
    \code{2}  & \code{2} & \code{1} & \(\hat{\theta}_{2,2,1}\) \\
    \code{3}  & \code{2} & \code{1} & \(\hat{\theta}_{3,2,1}\) \\
    \code{1}  & \code{1} & \code{2} & \(\hat{\theta}_{1,1,2}\) \\
    \code{2}  & \code{1} & \code{2} & \(\hat{\theta}_{2,1,2}\) \\
    \code{3}  & \code{1} & \code{2} & \(\hat{\theta}_{3,1,2}\) \\
    \code{1}  & \code{2} & \code{2} & \(\hat{\theta}_{1,2,2}\) \\
    \code{2}  & \code{2} & \code{2} & \(\hat{\theta}_{2,2,2}\) \\
    \code{3}  & \code{2} & \code{2} & \(\hat{\theta}_{3,2,2}\) \\
    \(\vdots\)  & \(\vdots\) & \(\vdots\) & \(\vdots\) \\
    \hline
  \end{tabular}
\end{table}

Once a dataset has been uploaded via one of the three methods outlined, the user will have to define the variables required by \pkg{INTEREST} and some optional variables, depending on the structure of the input dataset.
The names of each column (i.e. variable) from the uploaded dataset automatically populate a set of select-list inputs to assist the user.

The only variable required by \pkg{INTEREST} is a variable defining a point estimate from the simulation study; users can also pass standard errors of such estimates, and the true value of the estimand.
If neither of these values is provided, only performance measures that can actually be calculated with the available information are returned.
In order to provide additional flexibility, the user can define a column in the dataset that defines the true values of the estimand: this is especially useful e.g. in settings where the true value can vary between repetitions.
Further to that, a user can provide repetition-specific confidence bounds or even use t-distributed critical values rather than normal theory (by specifying a column that contains degrees of freedom per each repetition); once again, this can all be set via the \emph{Data} tab, and will affect relevant performance measures.
Finally, a user can define a variable representing methods being compared with the current simulation study (and choose the comparator), and one or more variables defining data-generating mechanisms (DGMs, e.g. sample size, true correlation, true baseline hazard function for survival models, etc.).
We denote with \emph{methods} the levels of the factor of primary comparative interest in a simulation study, and not necessarily an analytical method (strictly speaking).
Other factors e.g. characteristics of the data-generating mechanism can be used as well, if representing the primary comparative interest of a study.

In its current form, \pkg{INTEREST} can only accept a single column as a \emph{method} variable; when the primary focus of a simulation study is on several factors at once, we suggest pre-processing the dataset by creating a single column with all possible combinations from the factors of interest (e.g. using the \code{interaction} function in \proglang{R}).

The \emph{View uploaded data} side tab in \pkg{INTEREST} displays the dataset uploaded by the user using the \proglang{R} package \pkg{DT}, an \proglang{R} interface to the \pkg{DataTables} plug-in for jQuery \citep{dt}.
The resulting table is interactive and can be sorted and filtered by the user.
It is good practice to verify that the uploaded dataset is as expected before continuing with the analysis and any visual exploration.

\subsection{Missing data}

\pkg{INTEREST} includes a section for exploring missingness of estimates and/or standard errors from each repetition of a simulation study, which may occur, for example, due to non-convergence of some repetitions.
Missing values need to be carefully explored and handled at the initial stage of any analysis.
Missingness may originate as a consequence of software failures: if so, the code could (or should) be made more robust to ensure fewer or no failures.
Conversely, missing data may arise as a consequence of characteristics of the simulated data, yielding to non-convergence of the estimation procedures.
In other words, missing values may not be missing completely at random.
A discussion on the interpretation of missing values can be found elsewhere \citep{white_2011, morris_2019}.

The missing data functionality is based on the \proglang{R} package \pkg{naniar} \citep{naniar}, and can be accessed via the \emph{Missing data} tab.
It comprises visual and tabular summaries; missing data visualisations available in \pkg{INTEREST} are the following:
\begin{itemize}
    \item Bar plots of number (or proportion) of missing values by method and data-generating mechanism (if defined).
    Number and proportion of missing values are produced for each variable included in the data uploaded to \pkg{INTEREST};
    \item A plot to visualise the amount of missing data in the whole dataset;
    \item A scatter plot with missing status depicted with different colours; to be able to plot missing values, they are replaced with values 10\% lower than the minimum value in that variable.
    This plot allows identifying trends and patterns between variables in missing values (e.g. all estimates with a very large standard error have a missing point estimate);
    \item A heat plot with methods on the horizontal axis and the data-generating mechanisms on the vertical axis, with the colour fill representing the percentage of missingness in each tile.
\end{itemize}
Each plot can be further customised and exported (e.g. for use in slides and reports): more details in the \emph{Plots} section below.
Finally, \pkg{INTEREST} computes and outputs a table with the number, proportion, and the cumulative number of missing values per variable, stratifying by method and data-generating mechanisms; the table can be easily exported to \proglang{\LaTeX{}} format for further use (via the \code{kable} function from the \proglang{R} package \pkg{knitr} \citep{knitr}).

\subsection{Performance measures}

\pkg{INTEREST} estimates performance measures automatically as soon as the user defines the required variables via the \emph{Data} tab.
Supported performance measures are presented in Table \ref{tab:sstat}, and discussed in more detail elsewhere \citep{burton_2006, white_2010, morris_2019}.
In addition to that, \pkg{INTEREST} returns mean and median estimate, and mean and median squared error of the estimate.
Finally, \pkg{INTEREST} computes and returns Monte Carlo standard errors by default.
The list of performance measures estimated by \pkg{INTEREST} can be customised via the \emph{Options} tab: by default, all are included.

\begin{table}
  \caption{Overview of performance measures estimated by \pkg{INTEREST}.}
  \label{tab:sstat}
  \centering
  \begin{tabular}{rp{0.45\textwidth}}
    \toprule
    Performance measure                    & Description \\
    \midrule
    Bias                                   & Deviation between estimate and the true value \\
    Empirical standard error               & Log-run standard deviation of the estimator \\
    Relative precision against a reference & Precision of a method B compared to a reference method A \\
    Mean squared error                     & The sum of squared bias and variance of the estimator \\
    Model standard error                   & Average estimated standard error \\
    Coverage                               & Probability that a confidence interval contains the true value \\
    Bias-eliminated coverage               & Coverage after removing bias, i.e. by computing the probability that a confidence interval contains the average point estimate across repetitions instead of the true value \\
    Power                                  & Power of a significance test \\
    \bottomrule
  \end{tabular}
\end{table}

\subsection{Tables}

Estimated performance measures are presented in tabular form in the \emph{Performance measures} side tab, once again using the \proglang{R} package \pkg{DT}.
The table of estimated performance measures is relative to a given data-generating mechanism, which can be modified using a select list input on the side.
It is also possible to customise the number of significant digits and to select whether Monte Carlo standard errors should be excluded in each table or not via the \emph{Options} tab.

Finally, it is possible to export the tables in two ways:
\begin{enumerate}
  \item Export the table in \proglang{\LaTeX{}} format, e.g. for use in reports, articles, or presentations, via the \emph{Export table} tab and the \code{kable} function from the \proglang{R} package \pkg{knitr} \citep{knitr}.
  The caption of the table can be directly customised;
  \item Export estimated performance measures as a dataset, e.g. to be used with a different software package of choice.
  The table of estimated performance measures can be exported as displayed by \pkg{INTEREST} or in tidy format, and in a variety of formats: comma-separated (\code{.csv}), tab-separated (\code{.tsv}), \proglang{R} (\code{.rds}), \proglang{Stata} (version 8-15, \code{.dta}), \proglang{SPSS} (\code{.sav}), and \proglang{SAS} (\code{.sas7bdat}).
\end{enumerate}

\subsection{Plots}

\pkg{INTEREST} can produce a variety of plots to automatically visualise results from simulation studies.
Plots produced by \pkg{INTEREST} can be categorised into two broad groups: plots of estimates (and their estimated standard errors) and plots of performance, following analysis.
Plots for method-wise comparisons of estimated values and standard errors are:
\begin{itemize}
    \item Scatter plots;
    \item Bland-Altman plots \citep{altman_1983, bland_1999};
    \item Ridgeline plots \citep{ggridges};
    \item Contour and hexbin plots (as implemented in \pkg{ggplot2}'s \code{geom\_density\_2d} and \code{geom\_hex} geometric objects).
\end{itemize}
Each plot will include all data-generating mechanisms by default and allows comparing serial trends and the relative performance of methods included in the simulation study; contour and hexbin plots are especially useful to deal with overplotting.

Conversely, the following plots are supported for estimated performance:

\begin{itemize}
    \item Plots of performance measures with confidence intervals based on Monte Carlo standard errors.
    There are two variations of this plot: forest plots, and lolly plots.
    Both methods display the estimated performance measure alongside confidence intervals based on Monte Carlo standard errors; different methods are arranged side by side, either on the horizontal or on the vertical axis;
    \item Heat plots of performance measures: these plots are mosaic plots where the several methods being compared (if defined) are on the horizontal axis and the data-generating mechanisms are on the vertical axis.
    Then, each tile of the mosaic plot is coloured according to the value of a given performance measure.
    To the best of our knowledge, this is a novel way of visualising results from simulation studies, with an application in practice that can be found elsewhere \citep{gasparini_2019};
    \item Zip plots to visually explain coverage probabilities by plotting the confidence intervals directly.
		More information on zip plots is presented elsewhere \citep{morris_2019};
    \item Nested loop plots, useful to compare performance measures from studies with several DGMs at once.
    This visualisation is described in more detail elsewhere \citep{rucker_2014}.
\end{itemize}

Finally, all plots can be exported for use in manuscript, reports, or presentations by simply clicking the \emph{Save plot} button underneath a plot; all plots are exported by default in \code{.png} format, but other options are available via the \emph{Options} tab.
For instance, to suit a wide variety of possible use cases, \pkg{INTEREST} supports several alternative image formats such as \code{pdf}, \code{svg}, and \code{eps}.
Through the \emph{Options} tab it is also possible to customise the resolution of the plot for non-vectorial format (in dots per inch, \code{dpi}) and the physical size (height and width) of the plots to be exported.
The \emph{Options} tab allows further customisations: for instance, it is possible to (1) define a custom label for the x-axis and the y-axis and (2) change the overall appearance of the plot by applying one of the predefined themes (which are described in more detail in the \emph{User guide} tab).

\subsection{Interactive apps for exploring results}

\pkg{INTEREST} allows researchers to upload a dataset with the results of their Monte Carlo simulation study obtaining estimates of performance in a quick and straightforward way.
This is very appealing, especially with simulation studies with several data-generating mechanisms where it could be confusing to investigate all scenarios at once.
Using the app it is possible to vary data-generating mechanisms and obtain updated tables and plots in real-time, therefore allowing to quickly iterate and take into consideration all possible scenarios.

\subsection{Interactive apps for disseminating results}

One of the intended usage scenarios for \pkg{INTEREST} consists of supplementing reporting of simulation studies.
This is especially useful with large simulation studies, where it is most cumbersome to summarise all results in a manuscript: it is common to include in the main manuscript only a subset of results for conciseness.
The remaining results are then relegated to supplementary material, web appendices, or not published at all - undermining dissemination and replicability of a study.

Furthermore, given that it is becoming increasingly common to publish the code of simulation study, one could publish the dataset with the results alongside the code used to obtain it.
That dataset could then be uploaded to \pkg{INTEREST} by readers, who could then explore the full results of the study as they wish.
Given the ubiquity of web services like GitHub (\url{https://github.com}) and data-sharing repositories such as Zenodo (\url{https://zenodo.org/}), we encourage \pkg{INTEREST} users to publish online the full results of their simulation studies for other users to download and experiment with.

\section{Future developments}

Although \pkg{INTEREST} is fully functional in its current state, several future developments are being planned.
For instance, we aim to include support for multiple estimands at once as currently supported by \pkg{rsimsum} via the \code{multisimsum} function.
We also aim to improve the flexibility of \pkg{INTEREST} in terms of customisation (of tables and plots), e.g. by displaying the raw \proglang{R} code used to generate the plots behind the scenes.
Finally, we are considering adding additional interactive features to the app via \proglang{HTML} widgets, \pkg{D\textsuperscript{3}}, or other approaches; there are several \proglang{R} packages that allow incorporating interactive graphs into \pkg{Shiny} apps such as \pkg{htmlwidgets} \citep{htmlwidgets}, \pkg{plotly} \citep{plotly}, and \pkg{r2d3} \citep{r2d3}.

\section{Case study}

The case study included in this Section illustrates the use of \pkg{INTEREST} to analyse publicly available results of a simulation study.
In particular, we will be using the results from the worked illustrative example included in Morris \emph{et al}. \citep{morris_2019}.

The study dataset contains the results of a simulation study comparing three different methods for estimating the hazard ratio in a randomised trial with a time to event outcome.
In particular, the methods being compared are proportional hazards survival models of the kind:
\[
    h_i(t) = h_0(t) \exp(X_i \theta),
\]
where \(\theta\) is the log hazard ratio for the effect of a binary exposure (e.g. treatment).
This class of models requires an assumption regarding the shape of the baseline hazard function \(h_0(t)\): it can be assumed to follow a given parametric distribution, or it can be left unspecified (yielding therefore a Cox model).
The \emph{aim} of this simulation study consists of assessing the impact of such an assumption on the estimation of the log hazard ratio.

Morris \emph{et al}. consider two distinct \emph{data-generating mechanisms}, varying the baseline hazard function:

\begin{enumerate}
    \item An exponential baseline hazard with \(\lambda = 0.1\) (DGM = 1);
    \item A Weibull baseline hazard with \(\lambda = 0.1, \gamma = 1.5\) (DGM = 2).
\end{enumerate}

In both settings, data are simulated on 300 patients with a binary covariate (e.g. treatment) simulated using \(X_i \sim \operatorname{Bern}(0.5)\) - simple randomisation with an equal allocation ratio.
The log hazard ratio is set to be \(\theta = -0.50\); this is the true value of the \emph{estimand} of interest.

Three distinct \emph{methods} are fit to each simulated scenario: a parametric survival model that assumes an exponential baseline hazard, a parametric survival model that assumes a Weibull baseline hazard, and a Cox semi-parametric survival model.

Finally, the \emph{performance measures} of interest are bias, coverage, empirical and model-based standard errors.
Assuming that \(\text{Var}(\hat{\theta}) \le 0.04\), 1600 repetitions are run to ensure that the Monte Carlo standard error of bias (the key performance measure of interest) is lower than 0.005.

The dataset with the results of this simulation study is publicly available in \proglang{Stata} format, and can be downloaded from a GitHub repository at the following URL:

\url{https://github.com/tpmorris/simtutorial/raw/master/Stata/estimates.dta}

Within the dataset published on GitHub, the exponential, Weibull, and Cox models are coded as model 1, 2, and 3, respectively.

\begin{figure}
  \caption{App interface to load the dataset for the case study. \pkg{INTEREST} can import datasets that are available online by simply pasting a link to it; then, the required variables can be defined via a list of pre-populated select inputs.}
  \label{fig:load-data}
  \centering
  \includegraphics[width = 0.95 \textwidth]{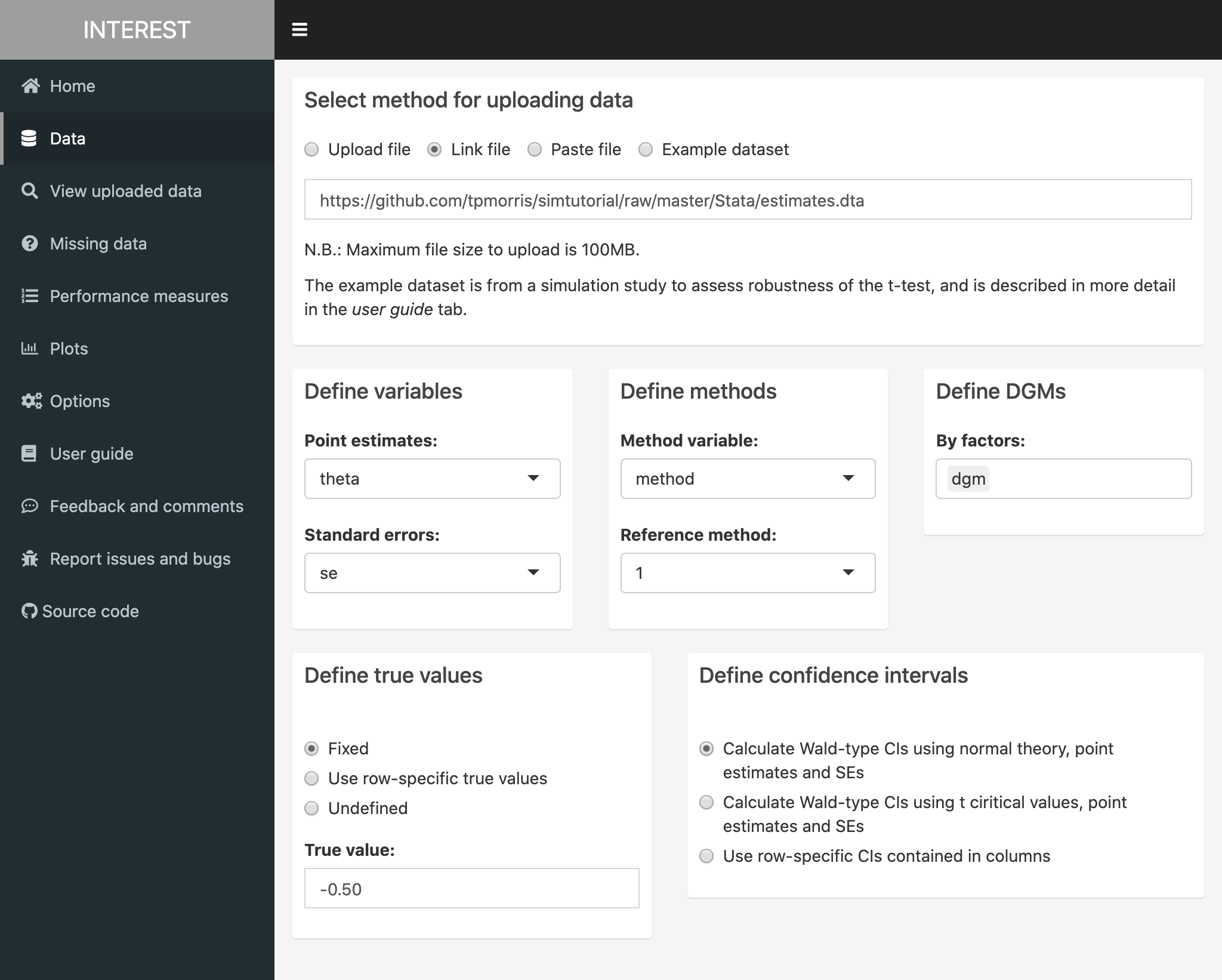}
\end{figure}

\begin{figure}
  \caption{Verifying the dataset for the case study. After importing the study dataset, it is recommended to verify that the uploaded data is correct.}
  \label{fig:view-data}
  \centering
  \includegraphics[width = 0.95 \textwidth]{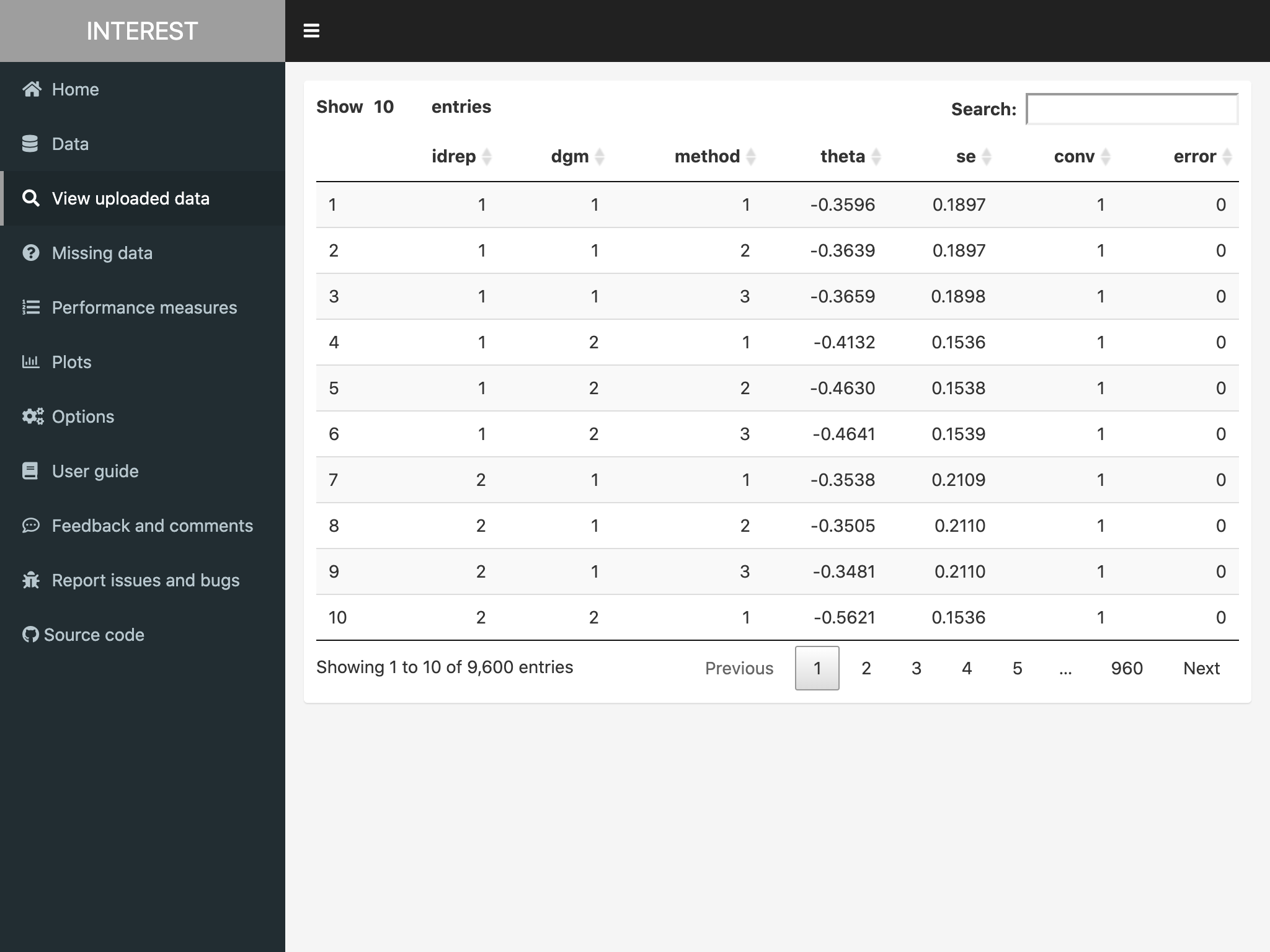}
\end{figure}

The workflow of \pkg{INTEREST} starts by providing the dataset with the results of the simulation study.
Given that the dataset is already available online, we can directly pass the URL above to \pkg{INTEREST} and then define the required variables (as illustrated in Figure \ref{fig:load-data}); the uploaded dataset can then be verified via the \emph{View uploaded data} tab (Figure \ref{fig:view-data}).

\begin{figure}
  \caption{Customising the performance measures reported by \pkg{INTEREST}. It is possible to focus on a subset of key performance measures by selecting them via the \emph{Options} tab.}
  \label{fig:custom-performance-measures}
  \centering
  \includegraphics[width = 0.95 \textwidth]{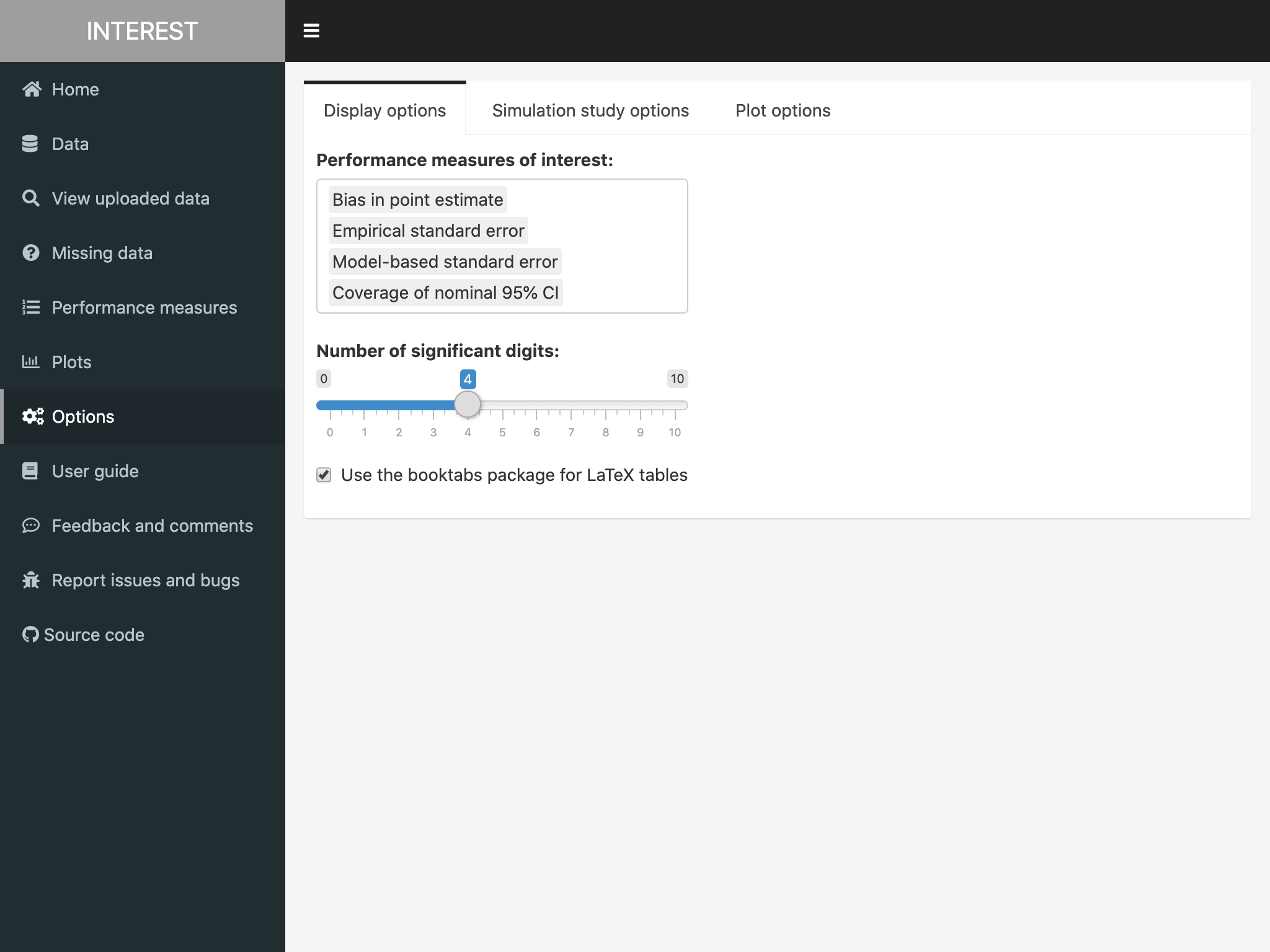}
\end{figure}

We can also customise the performance measures reported by \pkg{INTEREST} via the \emph{Options} tab (Figure \ref{fig:custom-performance-measures}), e.g. focussing on those outlined above as key performance measures (bias, coverage probability, empirical standard errors, model-based standard errors).

\begin{figure}
  \caption{Investigating missing data. Missingness patterns in the study dataset need to be assessed before continuing with the analysis. Several visualisations and tabular displays are available from the \emph{Missing data} tab.}
  \label{fig:missing-data}
  \centering
  \includegraphics[width = 0.95 \textwidth]{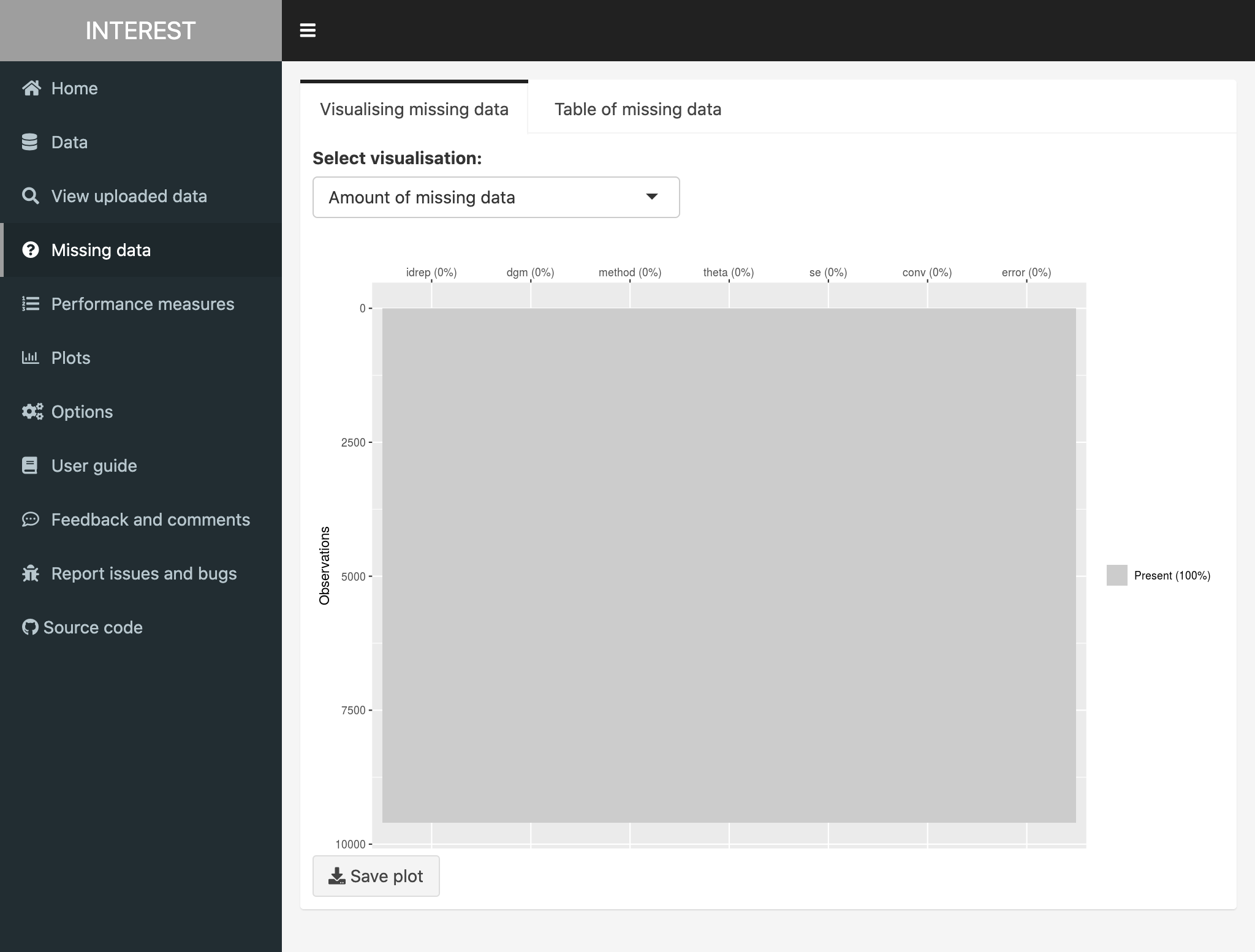}
\end{figure}

The next step of the workflow consists of investigating missing values: this can be achieved via the \emph{Missing data} tab.
In particular, there is no missing data in the study dataset (Figure \ref{fig:missing-data}).
We can, therefore, continue the analysis knowing that there is no pattern of serial missingness or non-convergence issues in our data.

\begin{figure}
  \caption{Table of performance measures for a given DGM. Performance measures of interest are tabulated in the \emph{Performance measures} tab, e.g. for the 2\textsuperscript{nd} DGM (with a Weibull baseline hazard function).}
  \label{fig:performance-measures-table}
  \centering
  \includegraphics[width = 0.95 \textwidth]{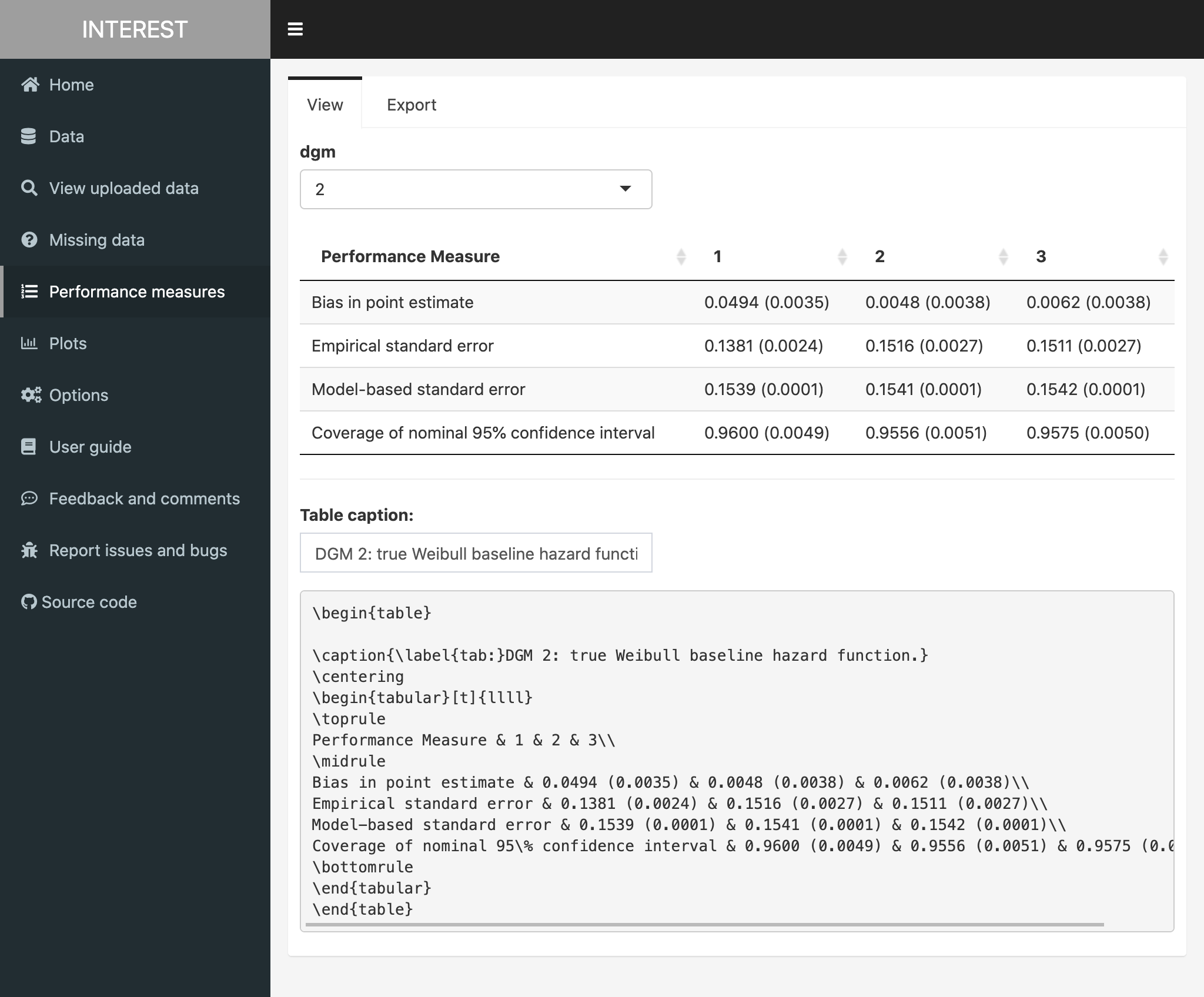}
\end{figure}

The performance measures of interest are tabulated in the \emph{Performance measures} tab, e.g. for DGM = 2 (Figure \ref{fig:performance-measures-table}).
We can see that bias for the exponential model is much larger than the Weibull and Cox models: approximately 10\% of the true value (in absolute terms) compared to less than 1\%.
Empirical and model-based standard errors are quite similar for the Weibull and Cox models; conversely, the exponential model seemed to overestimate the model-based standard error.
Coverage was as advertised for all methods, at approximately 95\%.
By comparison, all models performed equally in the other scenario (DGM = 1); these results are omitted from the manuscript for brevity, but we encourage readers to replicate this analysis and verify our statement.

\begin{figure}
  \caption{Exporting options for estimated performance measures. Performance measures of interest can be exported in a variety of formats ready to be used elsewhere (e.g. for dissemination purposes or to develop ad-hoc visualisations).}
  \label{fig:export-summary}
  \centering
  \includegraphics[width = 0.95 \textwidth]{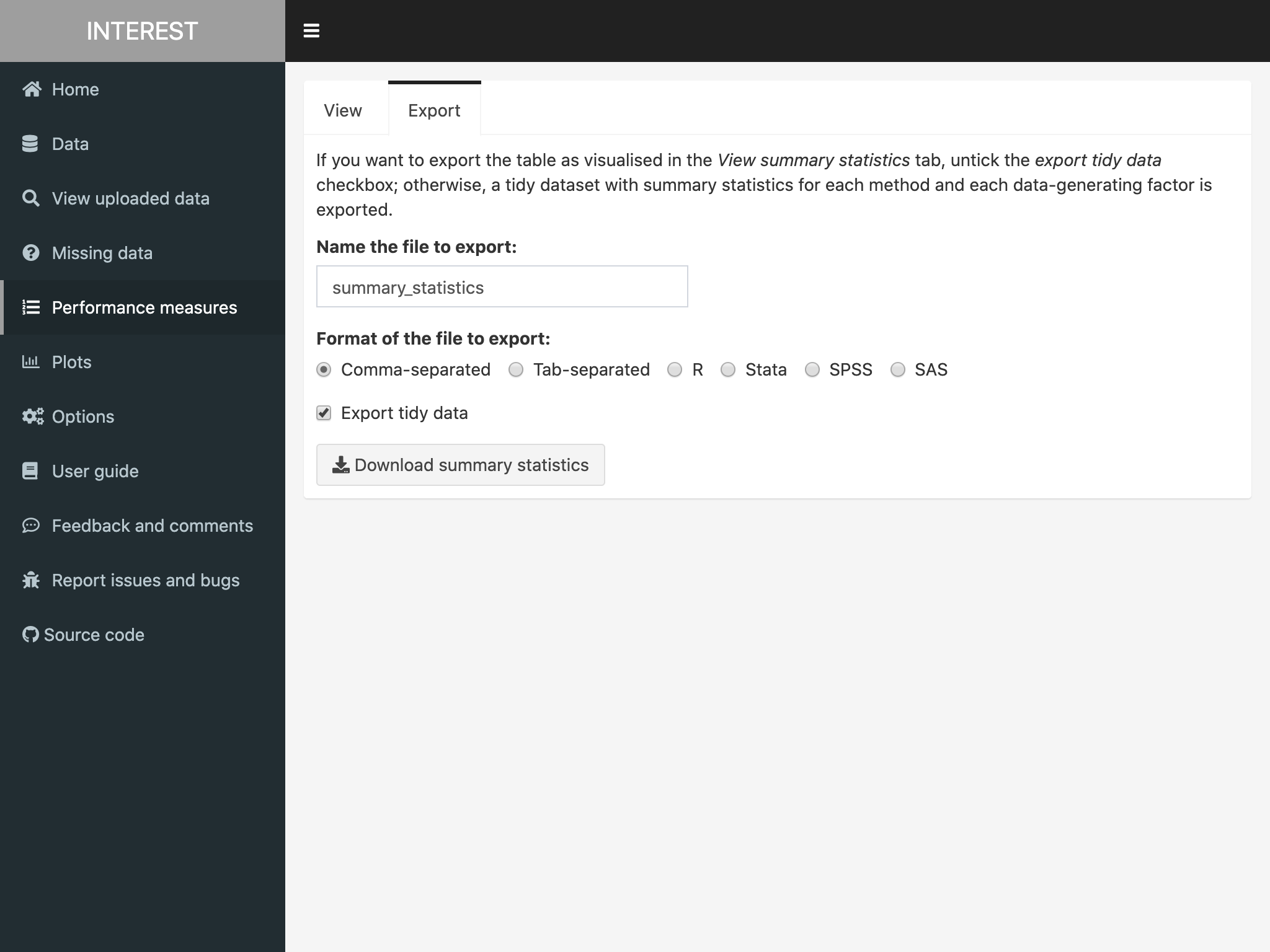}
\end{figure}

The \emph{Performance measures} tab provides a \LaTeX{} table ready to be pasted e.g. in a manuscript: the resulting table is included as Table \ref{tab:exported}.
A dataset with all the estimated performance measures here tabulated can also be exported to be used elsewhere (Figure \ref{fig:export-summary}).

\begin{figure}
  \caption{Visual comparison of point estimates via scatter plots. Points estimates for each method-DGM combination can be produced automatically using \pkg{INTEREST}.}
  \label{fig:est-vs-est}
  \centering
  \includegraphics[width = 0.95 \textwidth]{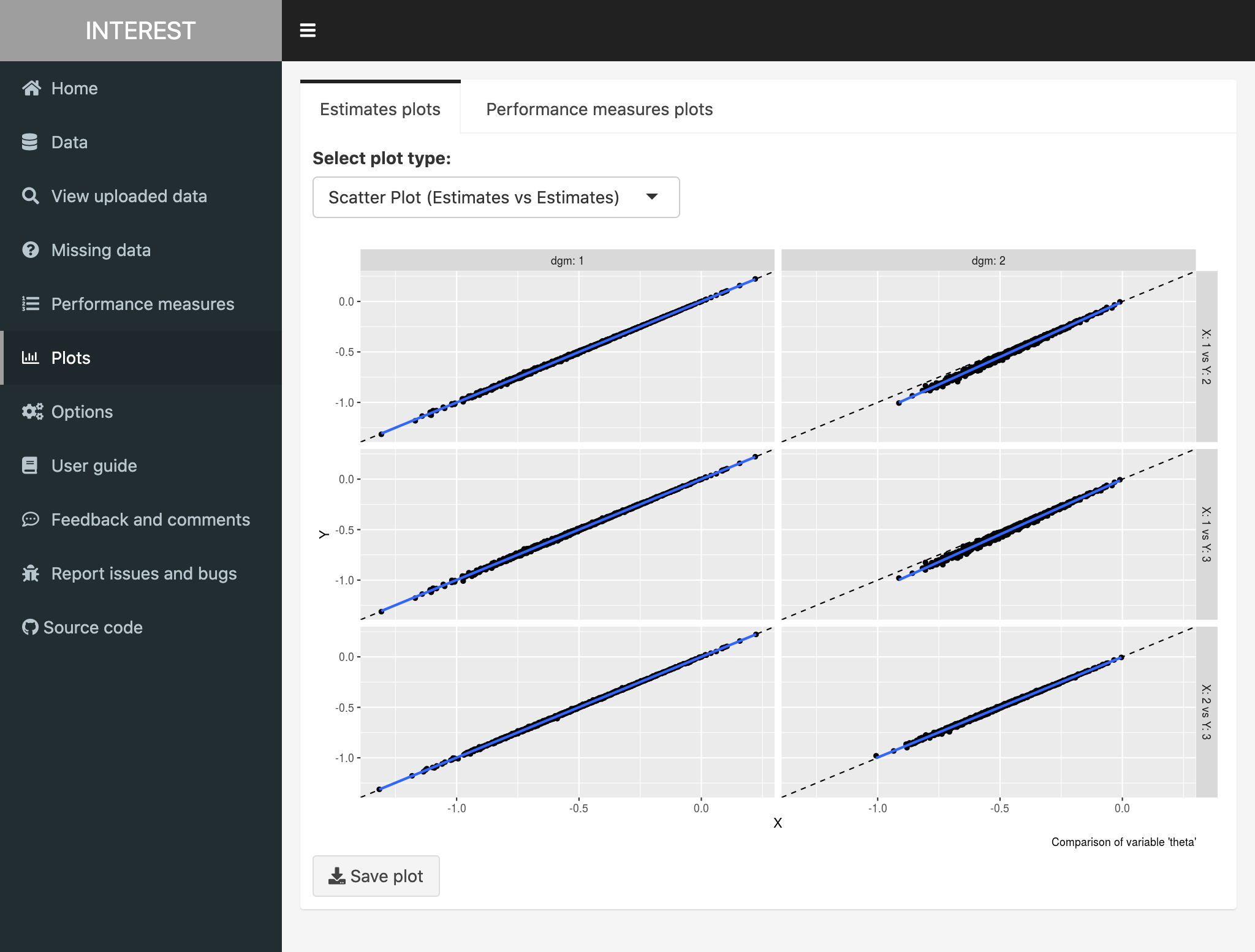}
\end{figure}

\begin{figure}
  \caption{Visual comparison of point estimates via Bland-Altman plots. Points estimates for each method-DGM combination can be produced automatically using \pkg{INTEREST}.}
  \label{fig:est-vs-est-ba}
  \centering
  \includegraphics[width = 0.95 \textwidth]{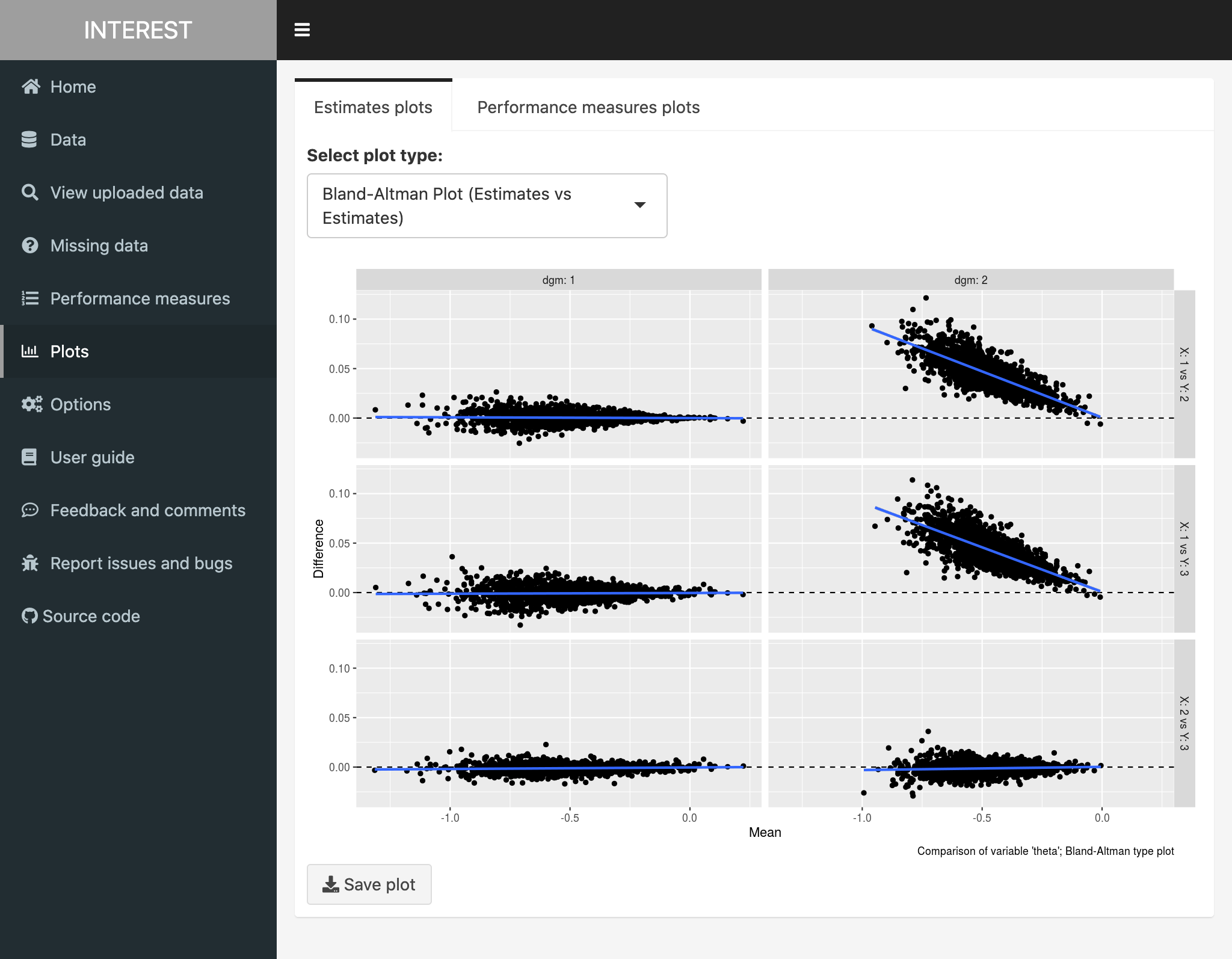}
\end{figure}

We can also visualise the results of this simulation study.
First, we can produce a method-wise comparison of point estimates from each method using e.g. scatter plots (Figure \ref{fig:est-vs-est}) or Bland-Altman plots (Figure \ref{fig:est-vs-est-ba}).
With both plots, it is possible to appreciate that for the DGM with \(\gamma = 1.5\) the exponential model yields point estimates that are quite different compared to the Weibull and Cox models.
Analogous plots can be obtained for estimated standard errors.

\begin{figure}
  \caption{Visual comparison of performance measures via forest plots. Estimated performance measures such as bias can be easily plotted via the \emph{Plots} tab.}
  \label{fig:bias-plot}
  \centering
  \includegraphics[width = 0.95 \textwidth]{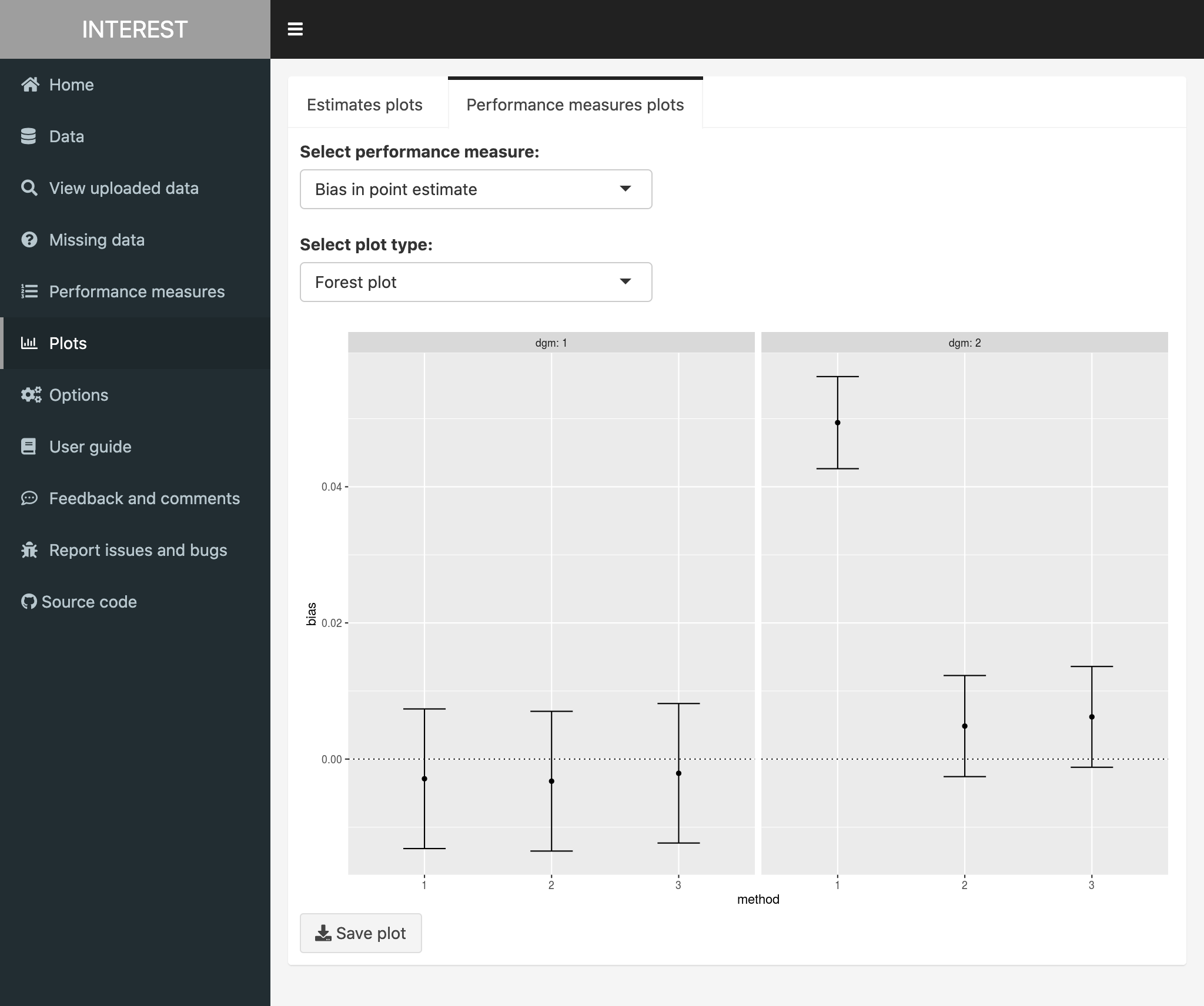}
\end{figure}

\begin{figure}
  \caption{Customising the visual appearance of plots. \pkg{INTEREST} allows customising the appearance of plots produced by the app via the \emph{Options} tab, e.g. by modifying the axes' labels and/or the overall theme.}
  \label{fig:plot-options}
  \centering
  \includegraphics[width = 0.95 \textwidth]{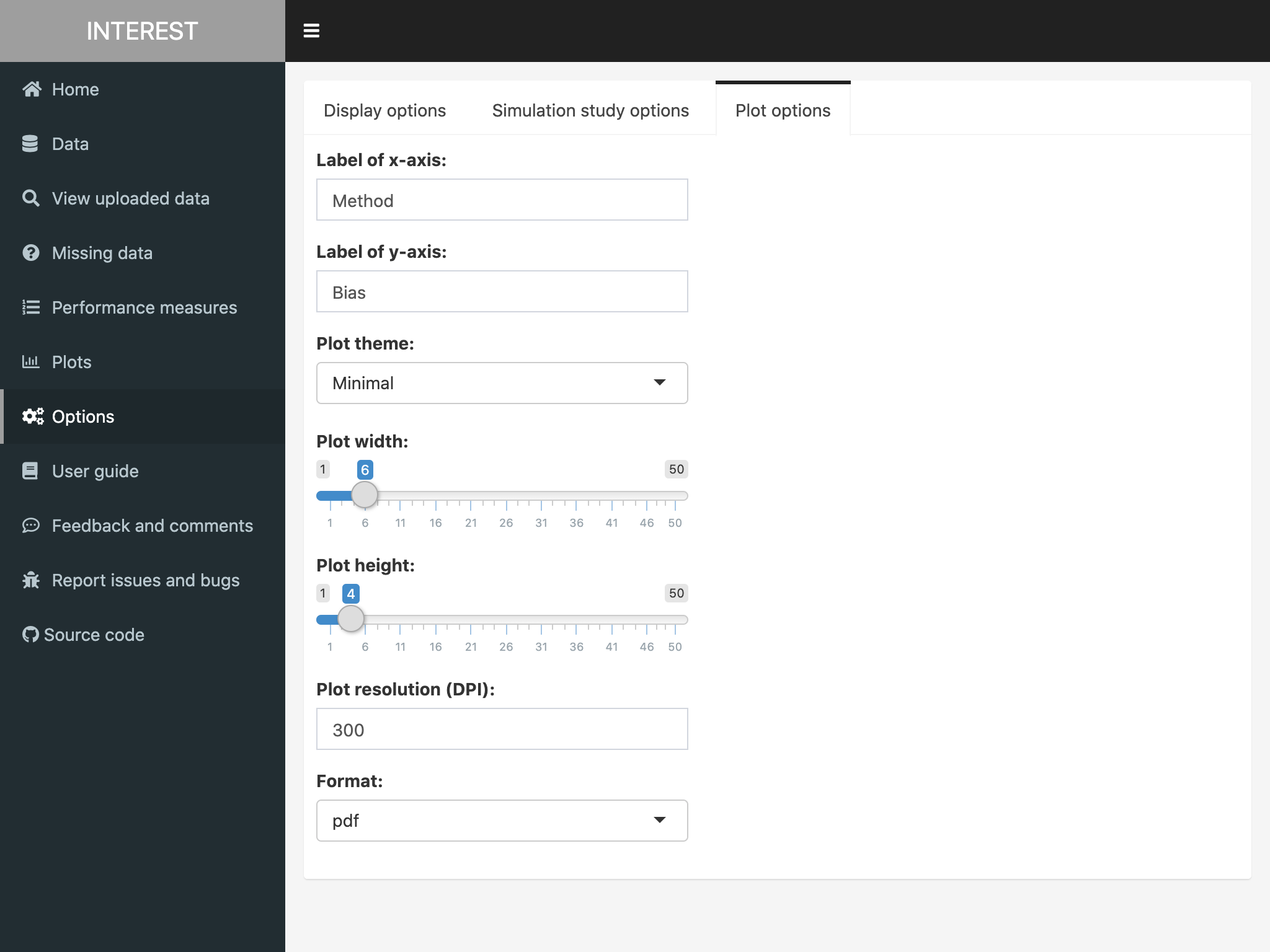}
\end{figure}

\begin{figure}
  \caption{Forest plot for bias, case study on survival regression modelling. This forest plot produced by \pkg{INTEREST} and further customised via the \emph{Options} tab can be directly exported from the app.}
  \label{fig:bias-plot-exported}
  \centering
  \includegraphics[width = 0.95 \textwidth]{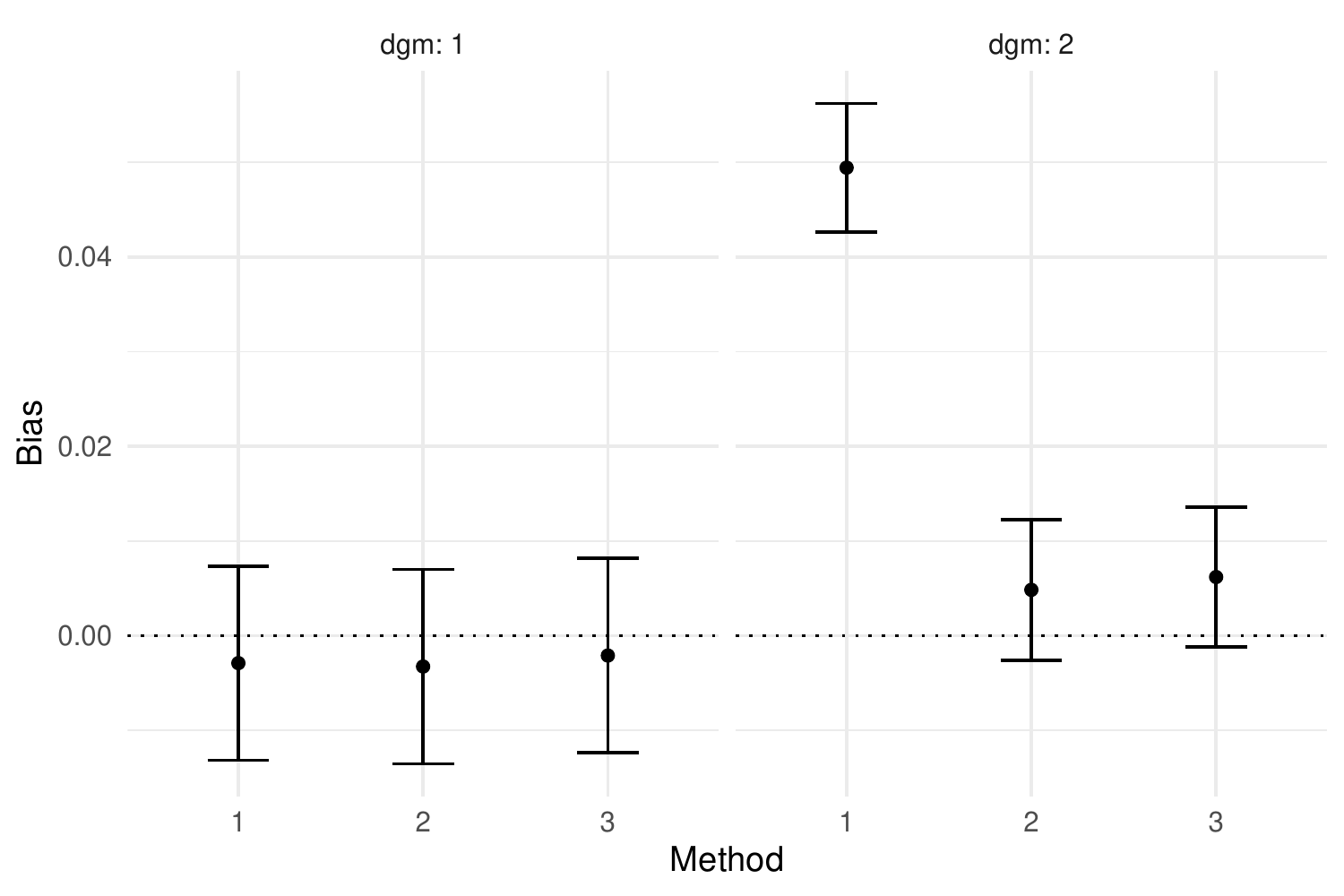}
\end{figure}

The performance measures tabulated in the \emph{Performance measures} tab can also be plotted via the \emph{Plots} tab.
For instance, it is straightforward to obtain a forest plot for bias (as illustrated in Figure \ref{fig:bias-plot}) which can be exported by clicking the \emph{Save plot} button.
The plots' appearance can also be customised via the \emph{Options} tab, e.g. by modifying the axes' labels and the overall theme of the plot (Figure \ref{fig:plot-options}); the resulting forest plot, exported in \code{.pdf} format, is included as Figure \ref{fig:bias-plot-exported}.
Several other data visualisations are supported by \pkg{INTEREST}, as described in the previous Sections: lolly plots, zip plots, and so on.

\begin{table}
  \centering
  \caption{Example of \LaTeX{} table directly exported from \pkg{INTEREST}, case study DGM 2: true Weibull baseline hazard function.}
  \label{tab:exported}
  \resizebox{\linewidth}{!}{
  \begin{tabular}[t]{llll}
  \toprule
  Performance Measure & 1 & 2 & 3\\
  \midrule
  Bias in point estimate & 0.0494 (0.0035) & 0.0048 (0.0038) & 0.0062 (0.0038)\\
  Empirical standard error & 0.1381 (0.0024) & 0.1516 (0.0027) & 0.1511 (0.0027)\\
  Model-based standard error & 0.1539 (0.0001) & 0.1541 (0.0001) & 0.1542 (0.0001)\\
  Coverage of nominal 95\% confidence interval & 0.9600 (0.0049) & 0.9556 (0.0051) & 0.9575 (0.0050)\\
  \bottomrule
  \end{tabular}
  }
  \end{table}

\section{Conclusions}

As outlined in the introduction, Monte Carlo simulation studies are too often poorly analysed and reported \citep{morris_2019}.
Given the increased use in methodological statistical research, we hope that \pkg{INTEREST} could improve reporting and disseminating results from simulation studies to a large extent.
As illustrated in the case study, the exploration and analysis of the Monte Carlo simulation study of Morris \emph{et al}. can be fully reproduced by using \pkg{INTEREST}.
Estimated performance measures are tabulated automatically, and plots can be used to visualise the performance measures of interest.
Moreover, the user is not constrained to a given set of plots and can fully explore the results with ease e.g. by varying DGMs to focus on or by choosing different data visualisations.
Most interestingly, the only requirement to reproduce the simulation study described in the case study is a device with a web browser and connection to the Internet.
To the best of our knowledge, there is no similar application readily available to be used by researchers and readers of published Monte Carlo simulation studies alike.


\section*{Acknowledgements}

TPM is supported by the Medical Research Council (grant numbers MC\_UU\_12023/21 and MC\_UU\_12023/29).
MJC is partially funded by the MRC-NIHR Methodology Research Panel (MR/P015433/1).

We thank Ian R. White for discussions that lead to the inception and development of \pkg{INTEREST}.


\bibliography{interest}

\end{document}